\documentclass[sigconf]{acmart}
\AtBeginDocument{%
  }

\copyrightyear{2026}
\acmYear{2026}
\setcopyright{cc}
\setcctype{by}
\acmConference[FSE Companion '26]{34th ACM Joint European Software Engineering Conference and Symposium on the Foundations of Software Engineering}{July 05--09, 2026}{Montreal, QC, Canada}
\acmBooktitle{34th ACM Joint European Software Engineering Conference and Symposium on the Foundations of Software Engineering (FSE Companion '26), July 05--09, 2026, Montreal, QC, Canada}
\acmDOI{10.1145/3803437.3805574}
\acmISBN{979-8-4007-2636-1/2026/07}

\usepackage{alltt}
\usepackage{microtype}
\usepackage{graphicx}
\usepackage{subcaption}
\usepackage{booktabs} 
\usepackage{multirow}
\usepackage{soul}
\include{todonotes}
\usepackage{hyperref}
\usepackage{tikz}
\usepackage{enumitem}
\usepackage{balance}

\usepackage{xcolor}
\usepackage{natbib}
\usepackage{graphicx}
\usepackage{listings}
\usepackage{subcaption}

\definecolor{background}{HTML}{F7F7F7}
\definecolor{keyword}{HTML}{37AC4A} 
\definecolor{operator}{HTML}{A51DFF}
\definecolor{string}{HTML}{C03333}

\lstdefinelanguage{python}{
  stringstyle=\textcolor{string},
  showstringspaces=false,
  morestring=[b]",
  morestring=[b]',
  morestring=[b]""",
  morecomment=[l]\#,
  morekeywords={and,as,assert,break,class,continue,def,del,elif,else,except,False,finally,for,from,global,if,import,in,is,lambda,None,nonlocal,not,or,pass,raise,return,True,try,while,with,yield},
  keywordstyle=\color{keyword}\bf\sffamily,
  commentstyle=\color{gray}\bf\sffamily,
  literate=
    *{>>}{{\bf\texttt{\color{operator}{>{}>}}}}{1}
    {\&}{{\bf\texttt{\color{operator}{\&}}}}{1}
    {|}{{\bf\texttt{\color{operator}{|}}}}{1}
    {=}{{\bf\texttt{\color{operator}{=}}}}{1}
    {(}{{\bf\texttt{\color{keyword}{(}}}}{1}
    {)}{{\bf\texttt{\color{keyword}{)}}}}{1}
    {[}{{\bf\texttt{\color{keyword}{[}}}}{1}
    {]}{{\bf\texttt{\color{keyword}{]}}}}{1}
    {\{}{{\bf\texttt{\color{keyword}{\char '173}}}}{1}
    {\}}{{\bf\texttt{\color{keyword}{\char '175}}}}{1},
}

\usepackage{pifont}

\usepackage{cuted}
\definecolor{Gray}{gray}{0.3}
\tikzstyle{mybox} = [draw=black, very thick, rectangle, rounded corners, inner ysep=5pt, inner xsep=5pt, fill=gray!20]

\usepackage{xspace}

\newcommand{\cmark}{\textcolor{green!70!black}{\ding{51}}} 
\newcommand{\xmark}{\textcolor{red}{\ding{55}}} 
\usepackage[capitalize,noabbrev]{cleveref}

\begin{document}

\title{Investigating Test Overfitting on SWE-bench}

\author{Toufique Ahmed, Jatin Ganhotra, Avraham Shinnar, and Martin Hirzel}
\affiliation{%
  \institution{IBM}
  \city{Yorktown Heights}
  \state{New York}
  \country{USA}
  \postcode{10603}}
\email{tfahmed@ibm.com, {jatinganhotra, shinnar, hirzel}@us.ibm.com}

\begin{abstract}
Tests can be useful towards resolving issues on code repositories.
However, relying too much on tests for issue resolution can lead to
code that technically passes observed tests but actually misses
important cases or even breaks functionality.
This problem, called \emph{test overfitting}, is exacerbated by the
fact that issues usually lack readily executable tests.
Instead, several issue resolution systems use tests auto-generated
from issues, which may be imperfect.
Some systems even iteratively refine code and tests jointly.
This paper presents the first empirical study of test overfitting
in this setting.

\end{abstract}

\ccsdesc[500]{Software and its engineering~Automatic programming}

\keywords{LLMs, SWE Patches, Reproduction Tests}


\maketitle

\section{Introduction}\label{sec:intro}

Issue resolution systems, which automatically resolve bug reports or
feature requests in code repositories, are useful.
They are also challenging to build, because issue descriptions tend to
be informal and code repositories are often large and
complex~\cite{jimenez_et_al_2024}.
Several issue resolution systems leverage tests: for code candidate
selection in parallel inference scaling~\cite{arora_et_al_2024,gao_et_al_2025,jain_et_al_2025,ruan_zhang_roychoudhury_2024,xia_et_al_2025,zhang_et_al_2025},
code refinement in sequential inference scaling~\cite{chen_et_al_2024,li_et_al_2025_patchpilot,yang_et_al_2024,zhang_et_al_2024},
or both~\cite{ehrlich_et_al_2025,li_et_al_2025_infcode}.
These systems rely upon the implicit premise that tests offer a
reliable signal in this setting.
Unfortunately, little is known about whether this premise holds.
To the contrary, startling results in adjacent settings cast doubt upon
it~\cite{baker_et_al_2025,smith_et_al_2015,stroebl_kapoor_narayanan_2024}.
Relying upon tests for issue resolution can cause \emph{test
overfitting}: generated code that narrowly passes the observed tests
but breaks other functionality.
This paper empirically investigates the problem of test overfitting in
repository-level issue resolution, on benchmarks such as
SWE-bench~\cite{jimenez_et_al_2024}.

An issue resolution system takes an issue description~$d_\textrm{issue}$
along with the original code~$c_\textrm{old}$ of a repository as input and
produces modified code~$c_\textrm{new}$ as output.
While many repositories come with regression tests
\mbox{$t_\textrm{old}\subset c_\textrm{old}$}, those only test
existing behavior, not the desired functionality of~$d_\textrm{issue}$.
Instead, issue resolution systems are evaluated by running acceptance
tests~$t_\textrm{gold}$ on~$c_\textrm{new}$.
In SWE-bench, $t_\textrm{gold}$ is mined from issue-resolving pull
requests~\cite{jimenez_et_al_2024}.
Since most open issues lack acceptance tests, $t_\textrm{gold}$ is not yet
available during issue resolution --- neither in practice nor in the
benchmark.
Therefore, several issue resolution systems instead generate surrogate
tests~$t_\textrm{gen}$, based only on~$d_\textrm{issue}$
and~$c_\textrm{old}$.
Since generated tests~$t_\textrm{gen}$ are imperfect, over-relying on
them risks test overfitting, where~$c_\textrm{new}$
passes~$t_\textrm{gen}$ but fails the hidden~$t_\textrm{gold}$.

\begin{figure}
  \raggedright
  \underline{issue description $d_\textrm{issue}$}\\
  \begin{lstlisting}
JSONField values are displayed as dict when readonly in the admin.
For example, {"foo": "bar"} would be displayed as {'foo': 'bar'}.
  \end{lstlisting}
  \underline{code patch $c_\textrm{new}$ without refinement}
  \begin{lstlisting}[language=python]
+    elif isinstance(field, JSONField):
+       return field.prepare_value(value)
  \end{lstlisting}
  \underline{observed generated test $t_\textrm{gen}$}
  \begin{lstlisting}[language=python]
value, disp = {"foo": "bar"}, '{"foo": "bar"}'
self.assertEqual(
    display_for_field(value, JSONField(), empty), disp)
  \end{lstlisting}
  \underline{code patch $c'_\textrm{new}$ with refinement}
  \begin{lstlisting}[language=python]
+    elif isinstance(field, JSONField):
+        try:
+            return json.dumps(value)
+        except (TypeError, ValueError):
+            return empty_value_display
  \end{lstlisting}
  \underline{hidden golden test $t_\textrm{gold}$}
  \begin{lstlisting}[language=python]
for value, disp in [
    ({'a': {'b': 'c'}}, '{"a": {"b": "c"}}'),
    (['a', 'b'], '["a", "b"]'),
    ('a', '"a"'),
    ({('a', 'b'): 'c'}, "{('a', 'b'): 'c'}"),  # Invalid JSON.
]:
    self.assertEqual(
        display_for_field(value, JSONField(), empty), disp)
  \end{lstlisting}
  \vspace*{-2mm}
  \caption{\label{fig:example}Test overfitting example on django\_\_django-12308.}
  \vspace*{-2mm}
\end{figure}

Figure~\ref{fig:example} shows a concrete example.
Based on the issue description~$d_\textrm{issue}$, an issue resolution
system generates initial new code~$c_\textrm{new}$ and a reproduction test
generator generates a test~$t_\textrm{gen}$.
However, $c_\textrm{new}$ fails on $t_\textrm{gen}$, triggering
code refinement based on observed test~$t_\textrm{gen}$.
The refined code $c'_\textrm{new}$ passes $t_\textrm{gen}$ but
overfits to it, since it fails other cases covered by the hidden
golden test~$t_\textrm{gold}$.

This paper empirically investigates test overfitting in issue resolution
using popular large-scale open-source repositories and issues.
It was inspired by two earlier studies.
The paper ``Is the cure worse than the
disease?''~\cite{smith_et_al_2015} demonstrated that test overfitting
is a serious problem.
But unlike our paper, it predated large language models~(LLMs) and
focused on student projects.
The ``Inference scaling
{FLaws}''~\cite{stroebl_kapoor_narayanan_2024} paper found test
overfitting to be a limiting factor for inference scaling in
LLM-based code generation.
But unlike our paper, it did not study repository-level tasks, nor did
it explore test-based code refinement.
Overall, while these papers highlighted the problem in adjacent
settings, to our knowledge, ours is the first to explore it in the
setting of LLM-based issue resolution.

Our empirical study answers three research questions:
\begin{itemize}[nosep]
  \item[RQ1.] Do LLM-generated code patches $c_\textrm{new}$ overfit
    to tests $t_\textrm{gen}$ generated using the same LLM?
  \item[RQ2.] How does refining code~$c_\textrm{new}$ based on
    observed tests~$t_\textrm{gen}$ affect test overfitting?
  \item[RQ3.] If the hidden tests~$t_\textrm{gold}$ were revealed,
    would refining code based on them break other functionality?
\end{itemize}
\noindent
Our experiments use two LLMs, Claude-3.7 Sonnet and GPT-4o.
We use strong existing LLM-based systems as a starting point,
obtaining the candidate code~$c_\textrm{new}$ from
Agentless~\cite{xia_et_al_2025} and the surrogate
test~$t_\textrm{gen}$ from \mbox{e-Otter++ \cite{ahmed_et_al_2026}}.
We extend these systems with a test-based code refinement
loop that is representative of those found in recent
work~\cite{ehrlich_et_al_2025,li_et_al_2025_infcode,li_et_al_2025_patchpilot,li_et_al_2025_sstar}.
The experiments are based on the TDD-bench
Verified~\cite{ahmed_et_al_2024} benchmark for test-driven
development, which is derived from SWE-bench~\cite{jimenez_et_al_2024}.
TDD-bench Verified comprises challenging real-world
Python issue resolution tasks.
This paper makes the following contributions.

\begin{itemize}
  \item It measures the degree of test overfitting by LLMs in
    reposi\-tory-level issue resolution tasks.
  \item It describes techniques to reduce overfitting and reports how
    they impact both overfitting and performance.
  \item It also includes a limit study of how much an oracle with
    access to the hidden golden tests could help performance.
\end{itemize}

This paper demonstrates that test overfitting is a clear problem in
modern LLM-based issue resolution systems.
And unfortunately, while attempts to mitigate this overfitting do work
to some extent, they also reduce effectiveness on the original task of
issue resolution.
Even if golden tests were available, refining code to pass them can
still break other functionality.
Overall, we caution against over-reliance on tests during code
generation and hope to inspire future work in reducing test
overfitting.

\section{Background and Related Work}

Test overfitting happens when an system generates code that passes
observed tests but fails held-out hidden tests~\cite{smith_et_al_2015}.
Test overfitting is similar to overfitting a model to its training
data, where the model performs worse on held-out test data; but note
that in this work, we use only pre-trained LLMs without fine-tuning.
Instead, test overfitting means that generated code~$c_\textrm{new}$
is too narrow to generalize to held-out tests~$t_\textrm{gold}$.
Stroebl et al.\ report that this happens when a loop repeatedly
generates code for the HumanEval or MBPP benchmarks until finding code
that passes observed tests~\cite{stroebl_kapoor_narayanan_2024}.
Unlike Stroebl et al., we experiment with repository-level benchmarks
and we also explore test-based code refinement loops.

SWE-bench is a benchmark where the input comprises an issue
description~$d_\textrm{issue}$ and the original code~$c_\textrm{old}$
of an entire repository before issue resolution and the output is
repaired code~$c_\textrm{new}$~\cite{jimenez_et_al_2024}.
Solutions are evaluated on whether~$c_\textrm{new}$ passes a held-out
test set~$t_\textrm{gold}$.
To emulate the real-world situation where issues do not come with
readily executable tests, the benchmark does not make those held-out
tests available to the issue resolution system.
SWE-bench Verified is a subset of 500 high-quality SWE-bench
instances, targeted by most issue resolution
systems~\cite{chowdhury_et_al_2024}.
SWT-bench~\cite{mundler_et_al_2024} and
TDD-Bench Verified~\cite{ahmed_et_al_2024}
are reproduction test benchmarks derived from
SWE-bench and SWE-Bench Verified.

Several issue resolution systems leverage generated
tests~$t_\textrm{gen}$.
CodeT generates many code candidates and many candidate tests,
then uses a dual execution agreement to pick a single output
code~\cite{chen_et_al_2023}.
Agentless generates up to 40 candidate codes for SWE-bench and
generates a reproduction test, then uses the test to filter the
candidate codes as part of code selection~\cite{xia_et_al_2025}.
S*~\cite{li_et_al_2025_sstar}, CodeMonkeys~\cite{ehrlich_et_al_2025},
and InfCode~\cite{li_et_al_2025_infcode} also generate both candidate
codes and candidate tests, followed by a loop where the LLM improves
the code based on the test and vice versa.
These papers might suffer from test overfitting but do not study it.

Test overfitting is related to \emph{reward hacking}, where a reward
is a reinforcement-learning~(RL) objective, and a hack finds a
loop-hole for improving the reward while circumventing its intention.
CURE uses RL to co-evolve an LLM to generate both code and white-box
tests, and might be susceptible to test overfitting and reward
hacking~\cite{wang_et_al_2025}.
Baker et al.\ found that RL to improve frontier LLMs for
repository-level coding tasks can cause the reward hack of disabling
tests instead of improving code~\cite{baker_et_al_2025}.

\section{Methodology}\label{sec:method}

We assume a simple three-step pipeline for issue resolution.

\begin{enumerate}[leftmargin=5mm]
  \item Run an off-the-shelf issue resolver
    (Agentless~\cite{xia_et_al_2025}) on $d_\textrm{issue}$ and
    original code $c_\textrm{old}$ to obtain initial candidate
    code~$c_\textrm{new}$.
  \item Run an off-the-shelf reproduction test generator
    (\mbox{e-Otter++~\cite{ahmed_et_al_2026}}) on
    $d_\textrm{issue}$ and $c_\textrm{old}$ to obtain an initial
    candidate test~$t_\textrm{gen}$.
  \item Run test-based code refinement (described below) on the
    initial $c_\textrm{new}$ and $t_\textrm{gen}$ to obtain final
    candidate code~$c'_\textrm{new}$.
\end{enumerate}

\noindent
After that, we can quantify test overfitting by checking which
code~($c_\textrm{new}$ or~$c'_\textrm{new}$) passes which
tests~($t_\textrm{gen}$, $t_\textrm{gold}$, or~$t_\textrm{old}$).

Since Agentless~\cite{xia_et_al_2025} and
\mbox{e-Otter++~\cite{ahmed_et_al_2026}} have been described in
prior work, here we only briefly discuss characteristics salient to
our paper.
First, both Agentless and e-Otter++ internally employ some parallel
inference scaling.
In our pipeline, we only take the single final result selected by each
system and feed it into our own test-based code refinement stage,
which can be viewed as a form of sequential inference scaling.
Second, both Agentless and e-Otter++ internally make use of regression
tests \mbox{$t_\textrm{old}\subset c_\textrm{old}$} from the original
code of the repository.
This is a realistic reflection of how a human software engineer might
consult regression tests when tackling an issue.
At the same time, it only yields limited benefit, since existing tests
$t_\textrm{old}$ tend not to test the open issue~$d_\textrm{issue}$.

\begin{figure}
\centerline{\includegraphics[width=.9\columnwidth]{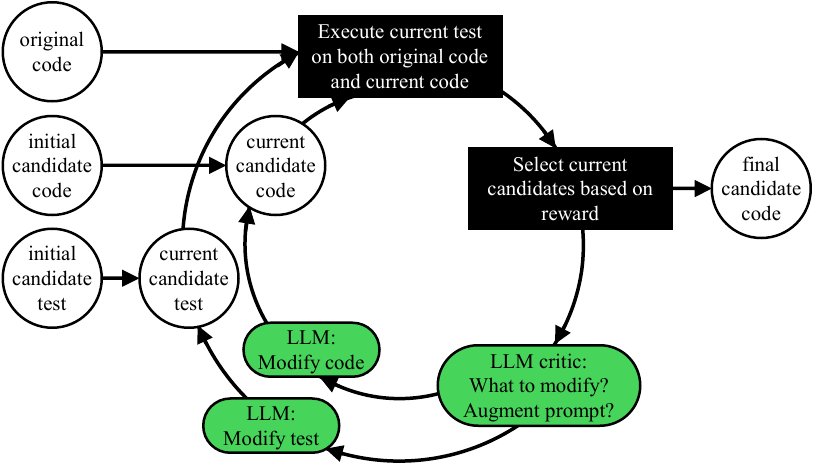}}
\caption{\label{fig:overview}Overview of test-based code refinement.}
\vspace{-12pt}
\end{figure}

Figure~\ref{fig:overview} shows our test-based code refinement approach. 
It starts with the initial candidate code~$c_\textrm{new}$ from Agentless
and the initial candidate test~$t_\textrm{gen}$ from \mbox{e-Otter++}.
We execute the test on $c_\textrm{old}$ and~$c_\textrm{new}$.
If the test goes from fail-to-pass, we immediately stop and declare success. 
Otherwise, we collect the execution logs from both test executions,
the focal functions modified by $c_\textrm{new}$, and the test
function~$t_\textrm{gen}$ and provide them to an LLM-based critic.
We ask the critic which part to modify~(focal or test). 
Inspired by \mbox{e-Otter++}, we also collect additional information such as the buggy line, relevant issue line, and lookup 
function (the function the model wants to see to fix it). Then, we ask the LLM to modify the focal code or test, 
exposing the counterpart~(e.g., for focal, the test) and additional information.
After updating the code~$c_\textrm{new}$ or test~$t_\textrm{gen}$, 
if the test is still not fail-to-pass, 
we need to decide whether to stick with our initial code or test or update it with the recently generated code or test, even though the test is not fail-to-pass.
To do this, we use a reward function to guide our decision:
$\mathit{Reward}(c_\mathit{old},c_\mathit{new},t_\textrm{gen})\,=$
\[ \frac{1}{3}\mathit{isFail}(t, c_\mathrm{old})+\frac{1}{3}\mathit{isPass}(t, c_\mathrm{new})+\frac{1}{3}\mathit{coverage}(c_\mathit{old},c_\mathit{new},t_\textrm{gen})
\] 
In the reward function, the first two components are binary, but coverage can vary from 0 to~1.
Since we know both the code patch and the test, we can compute how many deleted lines in $c_\mathrm{old}$ and how many added lines in $c_\mathrm{new}$ have been covered by the test. 
We divide the covered lines by the total number of added or deleted lines to compute coverage. 
If our new \mbox{$\langle c_\mathit{new},t_\textrm{gen}\rangle$} pair achieves
a higher reward score than the previous one, we replace it with the new one.
We repeat the improvement loop for at most 15 iterations if the observed test~$t_\textrm{gen}$ does not change from fail to pass.

\section{Results}


The experiments in this paper are based on Agentless\footnote{https://github.com/OpenAutoCoder/Agentless}-generated code patches and \mbox{e-Otter++}\footnote{https://zenodo.org/records/16755224}-generated tests.
Agentless was evaluated on SWE-bench Verified, and e-Otter++ was evaluated on TDD-Bench Verified, a subset of 449 instances from SWE-bench Verified. We evaluated on all 449 instances.
Our primary metric is test \emph{overfitting rate}.
A code patch $c_\textrm{new}$ is overfitting tests if it passes
$t_\textrm{gen}$ but fails $t_\textrm{gold}$ or~$t_\textrm{old}$.
Dividing the number of instances where that happens by the total instances yields the overfitting rate.

\begin{figure}
\centerline{\includegraphics[width=.9\columnwidth]{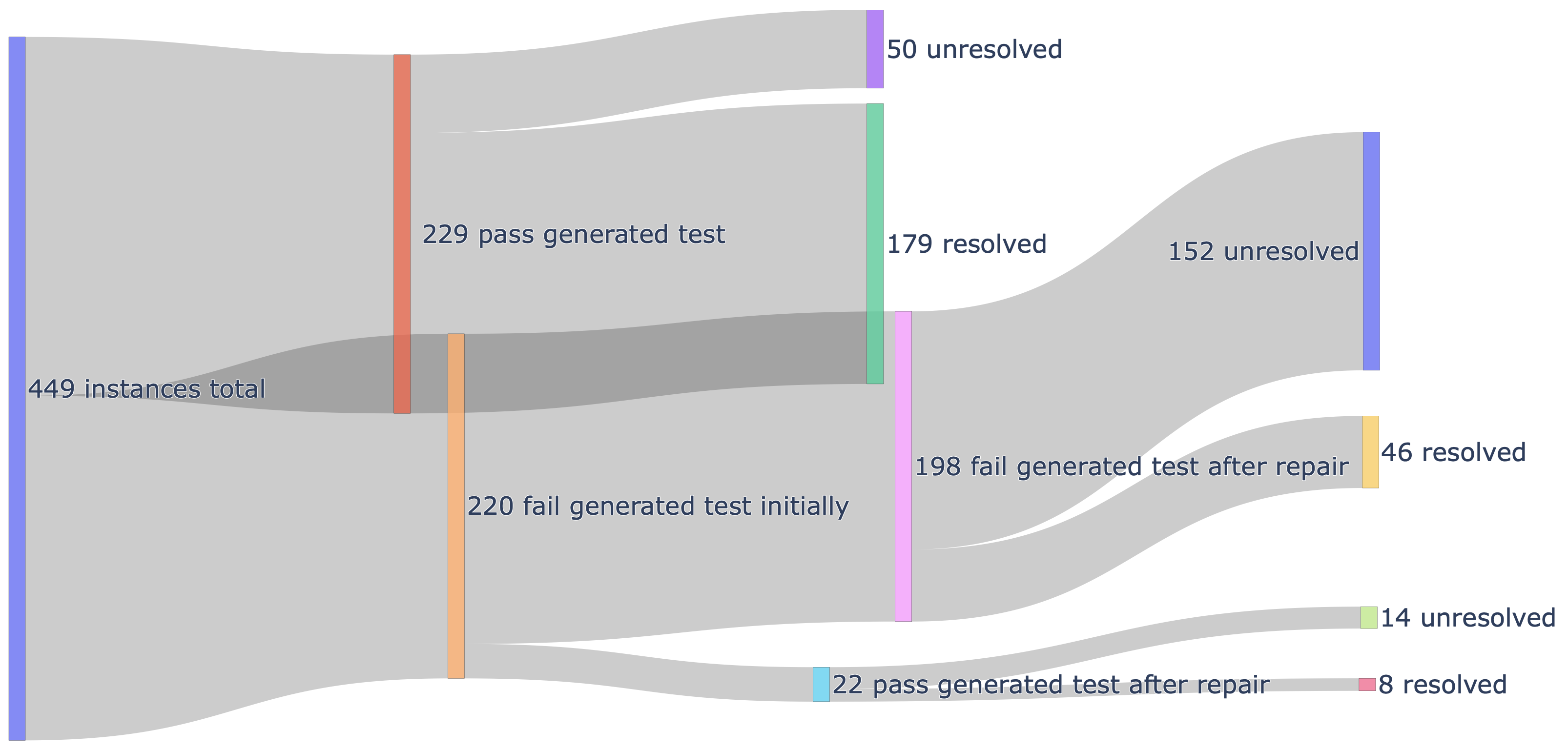}}
\caption{\label{fig:sankey}How severe is test overfitting? (Claude-3.7-Sonnet)}
\vspace{-5pt}
\end{figure}

\begin{table}[t]
\centering
\caption{Test overfitting rate of automated issue resolution.}
\resizebox{.95\columnwidth}{!}{%
\renewcommand{\arraystretch}{1.2}
\begin{tabular}{llrrrr}
\toprule    
\multicolumn{1}{c}{Model}          & \multicolumn{1}{c}{Setup} & \multicolumn{1}{c}{\begin{tabular}[c]{@{}c@{}}\# of \\ Sample\end{tabular}} & \multicolumn{1}{c}{Resolved} & \multicolumn{1}{c}{Unresolved} & \multicolumn{1}{c}{\begin{tabular}[c]{@{}c@{}}\%of \\ Overfitting\end{tabular}} \\ \midrule
\multirow{2}{*}{Claude-3.7-Sonnet} & w/o Refinement            & 229                                                                         & 179                          & 50                             & 21.8                                                                            \\
                                   & w/ Refinement             & 251                                                                         & 187                          & 64                             & 25.5                                                                            \\ \midrule
\multirow{2}{*}{GPT-4o}            & w/o Refinement            & 176                                                                         & 118                          & 58                             & 33.0                                                                            \\
                                   & w/ Refinement             & 198                                                                         & 127                          & 71                             & 35.9  \\ \bottomrule                                                                          
\end{tabular}
}
\label{tbl:rq1}
\vspace*{-2mm}
\end{table}

\subsection{Overfitting without Code Refinement (RQ1)}

To quantify overfitting without refinement, we run the e-Otter++ generated tests~$t_\textrm{gen}$ on Agentless code patches~$c_\textrm{new}$ and see how many pass.
Figure~\ref{fig:sankey} shows that with the Claude-3.7-Sonnet model, for 229 instances, the initial code passes the generated tests. Does this mean that the hidden golden tests~$t_\textrm{gold}$ will pass on these patches? 
The answer is no.
There are 50 instances where $t_\textrm{gold}$ failed, rendering them
unresolved according to the benchmark.
That means the overfitting rate among the 229 instances
\mbox{is $\frac{50}{229}=21.8\%$}.
For the GPT-4o model, the overfitting rate is 33.0\%, even higher than the Claude-3.7-Sonnet model (see Table~\ref{tbl:rq1}).
Note that a test can overfit due to weaknesses in both the code patches and the tests. However, we did not explicitly differentiate between the two, as doing so would require extensive qualitative studies, which are difficult to scale.



\subsection{Overfitting with Code Refinement (RQ2)}
Tests have been used to validate SWE-patches. What about using tests to improve patches at generation time?
We apply the test-based code refinement loop from
Section~\ref{sec:method} to each of the 220 instances from
Figure~\ref{fig:sankey} where the initial code~$c_\textrm{new}$ failed
the observed generated tests~$t_\textrm{gen}$.
This yielded 22 instances where the refined code~$c'_\textrm{new}$
passed the generated tests.
Unfortunately, 14 out of the 22 fail the hidden tests~$t_\textrm{gold}$, increasing the overfitting.
If we include these in our existing overfitted samples, the
overfitting rate increases to
\mbox{$\frac{50\,+\,14}{229+22}=\frac{64}{251}=25.5\%$}~(see Table~\ref{tbl:rq1}).
GPT-4o experiences a similar increase in overfitting, indicating that it is difficult to improve code patches by exposing tests to LLMs.

In the above experiments, the refinement loop exposed the generated
tests~$t_\textrm{gen}$ and additional information to the LLMs.
One idea to reduce overfitting might be to hide the tests and
additional information, so the LLMs may not be able to trick the tests.
Indeed, hiding the generated tests and revealing merely a pass/fail
flag to the LLM causes a drop in agreement between code patches and
generated tests.
However, the number of instances passing the hidden
tests~$t_\textrm{gold}$ also goes down a bit.
If we remove the additional information, it becomes more difficult for LLMs to overfit the patch. 
Unfortunately, we still see a high overfitting rate even after this intuitive overfitting reduction approach. 
The model can find a way by knowing that the existing test is not fail-to-pass (third row in Table~\ref{tbl:rq2}).
Note that this will not have any impact on instances that were already overfitted before the refinement.

\begin{table}[t]
\centering
\caption{Test overfitting rate of repaired samples, and effect over hiding information to mitigate overfitting.}
\resizebox{.95\columnwidth}{!}{%
\renewcommand{\arraystretch}{1.2}
\begin{tabular}{lccrrrr}
\toprule    
\multicolumn{1}{c}{Model}          & \begin{tabular}[c]{@{}c@{}}Expose \\ Test\end{tabular} & \begin{tabular}[c]{@{}c@{}}Augment \\ Prompt\end{tabular} & \multicolumn{1}{c}{\begin{tabular}[c]{@{}c@{}}\# of \\ Sample\end{tabular}} & \multicolumn{1}{c}{Resolved} & \multicolumn{1}{c}{Unresolved} & \multicolumn{1}{c}{\begin{tabular}[c]{@{}c@{}}\%of \\ Overfitting\end{tabular}} \\ \midrule
\multirow{3}{*}{Claude-3.7 Sonnet} & \cmark                                                      & \cmark                                                        & 22                                                                          & 8                            & 14                             & 63.6                                                                           \\
                                   & \xmark                                                      & \cmark                                                         & 18                                                                          & 5                            & 13                             & 72.2                                                                            \\
                                   & \xmark                                                      & \xmark                                                         & 14                                                                          & 6                            & 8                              & 57.1                                                                           \\ \midrule       
\multirow{3}{*}{GPT-4o}            & \cmark                                                     & \cmark                                                         & 22                                                                          & 9                          & 13                             & 59.1                                                                        \\
                                   & \xmark                                                     & \cmark                                                         & 17                                                                          & 5                           & 12                             & 70.6                                                                        \\
                                   & \xmark                                                      & \xmark                                                         & 16                                                                          & 5                           & 11                             & 68.8                                                                   \\ \bottomrule    
\end{tabular}
}
\label{tbl:rq2}
\vspace*{-2mm}
\end{table}

Table~\ref{tbl:rq2} shows that exposing tests and prompt augmentation resolved 8 samples. Do all these 8 samples help us resolve additional instances? 
The answer is no.
The initial code patch for one instance was already passing on the golden test.
And besides these 8, there were 2 instances that went from resolved to non-resolved.
So the net gain is +5 instead of +8. 
That indicates the actual gain from test-based refinement is less than the apparent gain.
The refinement reward's coverage term may structurally push patches toward fitting~$t_\textrm{gen}$'s execution path. However, it is also believed that tests with higher coverage are qualitatively better than other tests.

\subsection{Impact of Revealing Golden Tests (RQ3)}

RQ3 asks what would happen if the test-based code refinement would
not use imperfect generated tests~$t_\textrm{gen}$ but rather
golden tests~$t_\textrm{gold}$.
This is a limit study, because in practice, issues do not usually
come with oracular executable acceptance tests.
Table~\ref{tbl:rq3} shows how much overfitting occurs in this setting.
Here, the final determination of whether an instance is resolved is
based on passing both golden tests~$t_\textrm{gold}$ and a subset of
regression tests~$t_\textrm{old}$.
(While issue-resolution systems such as Agentless have access to the
full regression testing suite contained in the original
code~$c_\textrm{old}$, they do not know which subset of those tests
are expected to pass after issue resolution~\cite{chen_et_al_2026}.)
The overfitting rate, which means passing now-revealed golden
tests~$t_\textrm{gold}$ but failing the relevant subset of regression
tests~$t_\textrm{old}$, is low~(13/223=5.8\% and 22/194=11.3\% with
Claude and GPT-4o, respectively).
Apart from overfitting, this also shows how well the model performs
when golden tests~$t_\textrm{gold}$ are exposed.
When the model passes the golden reproduction tests, the probability
that the issue is resolved is high, but not 100\%, because regression
tests can still fail.
The improvement from using golden tests in issue resolution is
modest: the number of resolved instances with the Claude model goes
up from 229 to 243, around a 3\% increase.
This experiment emulates a situation where developers follow
test-driven development and write tests prior to
generating~$c_\textrm{new}$.
Unfortunately, the gains are limited, highlighting that current issue
resolution systems often miss the bigger picture.


\begin{table}[t]
\centering
\caption{Impact of revealing golden test on issue resolution.}
\resizebox{.95\columnwidth}{!}{%
\renewcommand{\arraystretch}{1.2}
\begin{tabular}{lrrrr}
\toprule    
\multicolumn{1}{c}{Model} & \multicolumn{1}{c}{\begin{tabular}[c]{@{}c@{}}\# of \\ Sample\end{tabular}} & \multicolumn{1}{c}{Resolved} & \multicolumn{1}{c}{Unresolved} & \multicolumn{1}{c}{\begin{tabular}[c]{@{}c@{}}\%of \\ Overfitting\end{tabular}} \\ \midrule
Claude-3.7 Sonnet         & 223                                                                         & 210                          & 13                             & 5.8                                                                             \\
GPT-4o                    & 194                                                                          & 172                           & 22                           & 11.3     \\ \bottomrule                                                                   
\end{tabular}
}
\label{tbl:rq3}
\vspace*{-3.5mm}
\end{table}

\section{Discussion}

\paragraph{LLMs' Preference on Focal and Test}

Test-based code refinement gives LLMs a choice to modify both the code~(focal function) and/or the test. 
Table~\ref{tbl:preference} shows that for instances where refinement succeeds, the model mostly elects to modify the focal function instead of the test. 
For each scenario, the number of test updates is low compared to focal updates. Also, there are few cases where the LLMs tries to fix both the focal function and the test. 
That indicates the model has a preference for focal modification and believes that the tests are mostly perfect.

\begin{table}[t]
\centering
\caption{LLMs preference for focal/test modifications}
\resizebox{.95\columnwidth}{!}{%
\renewcommand{\arraystretch}{1.2}
\begin{tabular}{lllrrrr}
\toprule    
\multicolumn{1}{c}{Model}          & \multicolumn{1}{c}{\begin{tabular}[c]{@{}c@{}}Expose \\ Test\end{tabular}} & \multicolumn{1}{c}{\begin{tabular}[c]{@{}c@{}}Prompt \\ Augment\end{tabular}} & \multicolumn{1}{c}{\# of Sample} & \multicolumn{1}{c}{\begin{tabular}[c]{@{}c@{}}Focal \\ Update\end{tabular}} & \multicolumn{1}{c}{\begin{tabular}[c]{@{}c@{}}Test \\ Update\end{tabular}} & \multicolumn{1}{c}{\begin{tabular}[c]{@{}c@{}}Both \\ Update\end{tabular}} \\ \midrule
\multirow{4}{*}{Claude-3.7 Sonnet} & \cmark                                                                          & \cmark                                                                             & 22                               & 18                                                                          & 3                                                                          & 1                                                                          \\
                                   & \xmark                                                                          & \cmark                                                                            & 18                               & 15                                                                          & 3                                                                          & 0                                                                          \\
                                   & \xmark                                                                          & \xmark                                                                            & 14                               & 13                                                                          & 0                                                                          & 1                                                                          \\
                                   & \cmark (Gold)                                                                    & \cmark (Gold)                                                                       & 25                               & 23                                                                          & 0                                                                          & 2                                                                          \\ \midrule  
\multirow{3}{*}{GPT-4o}            & \cmark                                                                          & \cmark                                                                             & 22                               & 16                                                                           & 2                                                                          & 4                                                                         \\
                                   & \xmark                                                                          & \cmark                                                                            & 17                               & 13                                                                          & 1                                                                         & 3                                                                        \\
                                   & \xmark                                                                          & \xmark                                                                             & 16                               & 11                                                                           & 1                                                                          & 4                                                                          \\
                                   & \cmark (Gold)                                                                    & \cmark (Gold)                                                                       & 28                               & 20                                                                          & 2                                                                         & 6       \\ \bottomrule                                                                  
\end{tabular}
}
\label{tbl:preference}
\vspace*{-2mm}
\end{table}


\paragraph{Coverage for Overfitted Sample}
We compare the coverage between unbiased code patches~(both~$t_\textrm{gen}$ and $t_\textrm{gold}$ pass) and 
overfitted patches for Claude-3.7.
The median for unbiased patches is~1, 
while for overfitted patches it is less than~0.8. 
Also, the mean coverage is higher for unbiased patches~(0.95 vs.\ 0.9). 
These results indicate that even though both unbiased and overfitted patches pass observed tests~$t_\textrm{gen}$, 
overfitted patches have lower coverage. 
Execution coverage or static analysis might be used as a warning against overfitting early during patch generation, without relying on hidden golden tests. We leave this for future work.

\section{Threats to Validity}
Since the benchmarks include only Python repositories, our findings may not generalize to other programming languages. 
We experimented with two models but did not investigate model contamination or memorization. 
However, prior works show that model-generated tests are very different from those written by developers. 
On another note, our approach may be considered a way to measure contamination by exposing overfitted patches. 
We leave this for future research. Other minor limitations include our reliance on the coverage package for reward formulation. 
The Python coverage package sometimes does not work due to dependency issues or multithreading. 
However, this happens in less than 1\% of cases, and we assign a coverage of 1 to mitigate the impact of imperfect coverage reports.
Our results can also be affected by test flakiness. However, we use pre-built docker containers for the experiments, which should minimize the impact of flakiness.

\section{Conclusion}
In this work, we systematically show that test overfitting is still a problem in LLM-based program repair. 
We also show that test-based refinement may resolve some additional issues but increases overfitting. 
Therefore, using tests for patch generation needs more deliberation.
We share the initial and modified code patches at this link: \url{https://doi.org/10.5281/zenodo.17227318}



\balance
\bibliographystyle{ACM-Reference-Format}
\bibliography{bibfile}

\end{document}